\documentclass{article} % For LaTeX2e
\usepackage{iclr2019_conference,times}

\usepackage{amsmath,amsfonts,bm}

\def\eqref#1{equation~\ref{#1}}
\def\1{\bm{1}}

\DeclareMathAlphabet{\mathsfit}{\encodingdefault}{\sfdefault}{m}{sl}
\SetMathAlphabet{\mathsfit}{bold}{\encodingdefault}{\sfdefault}{bx}{n}

\usepackage{hyperref}
\usepackage{url}
\usepackage{graphicx}
\usepackage{xcolor}
\usepackage{amsmath, amssymb}

\title{Robust posterior inference when statistically emulating forward simulations}

\author{Grigor Aslanyan, Richard Easther, Nathan Musoke, \& Layne C. Price  \\
Department of Physics \\
University of Auckland \\
Auckland, New Zealand
}

\iclrfinalcopy % Uncomment for camera-ready version, but NOT for submission.
\begin{document}

\maketitle

\begin{abstract}
Scientific analyses often rely on slow, but accurate forward models for observable data conditioned on known model parameters.  While various emulation schemes exist to approximate these slow calculations, these approaches are only safe if the approximations are well understood and controlled. This workshop submission reviews and updates a previously published method, which has been used in cosmological simulations, to (1) train an emulator while simultaneously estimating posterior probabilities with MCMC and (2) explicitly propagate the emulation error into errors on the posterior probabilities for model parameters.  We demonstrate how these techniques can be applied to quickly estimate posterior distributions for parameters of the  $\Lambda$CDM cosmology model, while also gauging the robustness of the emulator approximation.
\end{abstract}

\section{Introduction}\label{intro_sec}

In many scientific analyses, the core calculation performed is the determination of unknown model parameters $\theta$ from some empirical data $D$, typically utilizing a Bayesian framework to obtain a posterior probability distribution $P(\theta | D
)$ given prior probabilities on $\theta$ and a likelihood function $\mathcal L (D | \theta)$.  Frequently, the evaluation of $\mathcal L$  involves forward-simulating $D | \theta$ via computationally intensive procedures, such as solving a system of differential equations. A common approach to speeding up the expensive evaluation of $\mathcal L(D | \theta)$ is to train a secondary model $\mathcal E (D | \theta, \mathcal D_L)$, called the  \emph{emulator}, to approximate the output of $\mathcal L$ given a training set of pre-computed likelihood values at different points in parameter space: $\mathcal D_L \equiv \left \{\mathcal L(D | \theta_1), \mathcal L(D | \theta_2), \dots, \mathcal L(D | \theta_N) \right \}$.
 
While substituting the quick-to-evaluate $\mathcal E$ in place of the slow $\mathcal L$ in an analysis can reduce the computational time, the difference between the emulated and true likelihoods can introduce bias and variance into an estimated posterior distribution on $\theta$.  Also, the emulator $\mathcal E$ should be relatively easy to train, so that an effective reproduction of the likelihood function does not require dense sampling in $\mathcal D_L$, which defeats the purpose of training a separate emulator.  The literature on statistical emulators is well developed, with diverse field-specific applications; we refer the reader to \citet{kaufman2011efficient}, \citet{grow2014statistical}, and \citet{kasim2020up} for a detailed review of different techniques.  In this paper, we focus on robustness in emulation, instead of modeling details.

Here, we review a ``learn-as-you-go'' emulation algorithm, previously published by the authors~\citep{aslanyan2015learn} that dynamically trains an emulator and an error model for the differences between the emulated and exact likelihood functions.\footnote{Code available at \url{https://github.com/auckland-cosmo/LearnAsYouGoEmulator}. \nocite{layg}}
We use Markov Chain Monte Carlo (MCMC) sampling methods to  evaluate model posteriors and generate the training set for the emulator.  The local emulation errors, as estimated by the error model, are propagated through the calculation, resulting in an estimated error in the posterior probability $p(\theta | D)$.  Based on an externally defined threshold on the emulation error, we alternate between exactly evaluating $\mathcal L$ and approximating it, which allows us to trade off between accuracy and evaluation speed. Finally, we demonstrate that the algorithm can accelerate the calculation of posterior probabilities 67x-105x for parameters of interest in the standard, $\Lambda$CDM model of cosmology, without a pre-existing training set.

\section{Likelihood and posterior error modeling}\label{posterior_error_sec}
We approximate the likelihood function $\mathcal L (D | \theta)$ with an emulator function $\mathcal E(D | \theta, \mathcal D_L)$ and additionally learn an error model $e$ to estimate the difference between $\mathcal L$ and $\mathcal E$.  In this section, we discuss the details of these models.

\paragraph{Emulator training set, $\mathcal D_L$:} 
Given a set of $N$ points in parameter space $\Theta = \left\{ \theta_1, \dots, \theta_N \right \}$, where $\theta \in \mathbb{R}^m$, and a set of exactly evaluated likelihoods at these points $\mathcal D_L = \left\{\mathcal L(D|\theta_1), \dots, \mathcal L(D|\theta_N) \right \}$, we first calculate the covariance matrix $C$ of the $\theta_i$ in $\Theta$, then we decompose $C=L L^T$ into Cholesky matrices $L$.  We project $\Theta$ into the new basis, $\Theta^\prime = \left\{ L^{-1} \theta_i | \theta_i \in \Theta \right\}$, and construct a $k$-d tree $T_{\Theta^\prime}$ from $\Theta^\prime$.
The tree can be appended to as new elements enter the training set and periodically rebalanced; we found that a good rule of thumb is to rebalance $T_{\Theta^\prime}$ whenever the depth exceeds $4\log N$.

\paragraph{Emulator model, $\mathcal E$:} 
Any function approximation scheme can be used as an emulator in this approach.  Here, we describe a simple one that was effective in~\citet{aslanyan2015learn}, based on the Cholesky parameter-space projection $\Theta^\prime$ described above.  For a new point $\theta_0$ where we wish to emulate $\mathcal L(D | \theta_0)$, we first convert bases, $\theta_0^\prime = L^{-1} \theta_0$, then find the $k$ nearest parameter-space neighbors $\theta_0^\mathrm{NN}$ of $\theta_0^\prime$ in the $k$-d tree $T_{\Theta^\prime}$ and the corresponding exact likelihoods of these neighbors in $\mathcal D_L$. For each point in $\theta_0^\mathrm{NN}$ we assign a weight $w_i = 1/\mathrm{Dist}(\theta_0^\prime,\theta_i^\mathrm{NN})$, where the $\mathrm{Dist}$ function is a parameter-space metric; we choose simple $p$-norms.  Finally, we evaluate $\mathcal E(D | \theta_0)$ via polynomial interpolation, where the polynomial coefficient is found by weighted least-squares fit over the nearest neighbors and each neighbor is weighted by $w_i$.

\paragraph{Learned likelihood error model, $e$:}
In addition to training the emulator $\mathcal E$, we also estimate a transformed version of the local error 
$e(\theta) = \mathcal E(\theta) - \mathcal L(\theta)$.
We expect that the emulation error should scale monotonically with increasing distance between a newly sampled parameter point $\theta$ and its nearest points in the training set $\mathcal D_L$. Consequently, we assume $e \propto \rho(\theta | \mathcal D_L)$, where $\rho$ is a scalar that is inversely proportional to the mean $n$-dimensional Euclidean distance to the $k$ nearest neighbors in $\mathcal D_L$.
We then assume that $e$ increases linearly with $\rho$, allowing us to define a probability distribution on their ratio as
$
  p(e| \rho, \mathcal D_L)  \rightarrow p\left(e / \rho \, \big | \, \mathcal D_L^\mathrm{CV} \right),
$
up to an arbitrary normalization.  We estimate $p(e | \rho, \mathcal D_L)$ empirically, via cross-validation on subsets $\mathcal D_L^\mathrm{CV} \subset \mathcal D_L$, from which $e$ can be calculated exactly. 

\paragraph{Learn-as-you-go methods \& \emph{a priori} error threshold:}
In some cases, although evaluating $\mathcal L$ is slow, it is still feasible to sometimes exactly evaluate it. Exact evaluation is particularly useful in cases where the estimated emulation error $e$ is high.  We define "learn-as-you-go" methods as those situations where one attempts posterior estimation for parameters $\theta$ by first starting with an empty training set $\mathcal D_L$, iteratively adding elements to $\mathcal D_L$ by exactly sampling $\mathcal L$ at parameter space points as proposed by some external algorithm (such as MCMC), then periodically retraining both the emulator $\mathcal E$ and error model $e$; as $\mathcal E$ improves, it takes over all or part of the calculation.  We study this learn-as-you-go class in Section~\ref{sect:results}. 

Importantly, for this class of model, we can allow the user to define an error threshold $\epsilon$ such that if $e > \epsilon$, then we evaluate $\mathcal L$ exactly instead of emulating it, and add the new evaluation to the training set $\mathcal D_L$.  We use a threshold of 0.4 on the 68\% upper-bound of the distribution of errors on the quantity $-2 \log \mathcal L$, as estimated from the test set, as the default choice.

\paragraph{Inferred posterior error due to emulation:}
The output of our analysis will be posterior distributions, which can be evaluated at local points in parameter space $\theta_0 \in \mathbb R^m$, after marginalizing over nuisance parameters $\phi_N$:
%\begin{equation}\label{post_marg}
$ p(\theta_0 | D)=\int p(\theta_0, \phi_N |D) \, d\phi_N
  \;  = \; \frac{1}{Z}  \int \mathcal L(D|\theta_0, \phi_N) \, p(\theta_0, \phi_N) \, d \phi_N,
  $
%\end{equation}
where $Z(D) \equiv \int \mathcal{L}(D |\theta) p(\theta) d\theta$ is  the marginalized likelihood.  We are interested in how much error is introduced in the posterior distribution due to errors in the emulated likelihood function.  Importantly, we do not want to add any additional samples of the exact likelihood function in order to evaluate this error.

We express the error between the emulated posterior\footnote{We use  the subscript ``em'' on probability distributions achieved via emulation and no subscript on probability distributions that do not use emulation of the likelihood.} and the actual posterior,
$\Delta p(\theta_0 | D) \equiv p_\mathrm{em}(\theta_0 | D) - p(\theta_0 | D)$, in terms of emulated quantities as
\begin{equation}
  \Delta p (\theta_0 | D) = \frac{1}{Z_\mathrm{em}} \int p(\theta_0,\phi_N) \, \Delta \mathcal L (D | \theta_0, \phi_N) \, d \phi_N - \left[ \frac{p_\mathrm{em} (\phi_0 | D)}{Z_\mathrm{em}} \right]  \Delta Z(D) ,
  \label{eqn:Dp}
\end{equation}
where we have defined the emulation error $\Delta \mathcal L$ on the non-marginalized likelihood similarly to $\Delta p$ and the marginalized likelihood as
$\Delta Z(D) = \int p(\theta)  \, \Delta \mathcal L (D | \theta) \, d \theta$.  In~\eqref{eqn:Dp} we have assumed that the error terms $\Delta L$ and $\Delta Z$ are negligible at $\mathcal O(\Delta^2)$ and can be ignored. 

Finally, we define $\ell \equiv \log \mathcal L$ and rearrange~\eqref{eqn:Dp} to get the core result of our work:
\begin{equation}
  \label{eqn:post_error}
  \Delta \log p (\theta_0 | D) = \int \Delta \ell (D | \theta_0, \phi_N) \, \mathcal P (\phi_N | \theta_0) \, d \phi_N
                           - \int \Delta \ell (D | \theta, \phi_N)  \, p_\mathrm{em} (\theta, \phi_N | D) \, d \theta \, d\phi_N,
\end{equation}
where we have defined the normalized probability distribution
$\mathcal P (\phi_N | \theta_0) \equiv p_\mathrm{em}(\theta_0, \phi_N | D) \, / \, p_\mathrm{em}(\theta_0 | D).$
Equation~\ref{eqn:post_error} contains one term that arises from the emulated likelihood error locally around $\theta_0$ and a second term due to the overall posterior normalization.  The $\Delta \log \mathcal L$ term can then be estimated with our error model above. When using a user-defined error threshold $\epsilon$, any $\theta$ that would naively have an inferred error $e > \epsilon$ will instead be exactly evaluated with $\mathcal L$, yielding $\Delta \ell (D | \theta) \equiv 0$.

\paragraph{Estimating the inferred posterior error:}
We now need to approximate the integral in~\eqref{eqn:post_error}, which we do by histogram estimation.  We assume that we have a set of points $X_\mathrm{samp}=\left \{ (\theta, \phi_N)_\alpha \right \}_{\alpha=1}^{N_\mathrm{samp}}$ that are sampled from the emulated posterior $p_\mathrm{em}(\theta, \phi_N | D)$ via a Monte Carlo method.  For a given parameter point $\theta_0 \in \mathbb{R}^m$, we estimate the posterior on the $j^\mathrm{th}$ component of $\theta_0$, $p(\theta_{0,j} | D)$ by marginalizing over all other parameters $\left \{ \theta_{i \neq j}, \Phi_N \right \}_{i=1,m}$ in the sample, then using a histogram with $N_\mathrm{bin}$ bins, with appropriate normalization. Finally, we note that we can generate a sample $\Phi_\mathrm{samp}$ of nuisance parameters $\phi_N \sim p(\phi_N | \theta_{0,j})$  by collecting  the nuisance parameters of all posterior samples $\theta^\prime$ that have their $j^\mathrm{th}$ component in the same bin as $\theta_{0,j}$, since their posterior is the same.  This allows us to evaluate~\eqref{eqn:post_error} as:
\begin{equation}
  \Delta \log p(\theta_0 | D) \approx \frac{1}{N_\mathrm{bin}} \sum_{i=1}^{N_\mathrm{bin}} \Delta \ell (D | \theta_0, \phi_N^i )
  - \frac{1}{N_\mathrm{samp}} \sum_{j=1}^{N_\mathrm{samp}} \Delta \ell (D | \theta_j)
  \label{eqn:dp_sum}
\end{equation}
for $\theta_j \in X_\mathrm{samp}$ and $\phi_N^i \in \Phi_\mathrm{samp}$.  Equation~\ref{eqn:dp_sum} is written with an approximation symbol due to the histogram density estimator assumption, as well as replacing $\Delta \ell$ with the learned error model.
To gain intuition on \eqref{eqn:dp_sum}, we can further simplify if we assume that the emulation errors are approximately uncorrelated for different $\theta$, and $N_\mathrm{bin}$ and $N$ are large enough to invoke the Lyapunov central limit theorem.  Here, the distribution of log-posterior error terms will be approximately normally distributed with mean 
$\bar \mu_0 \equiv \frac{1}{N_\mathrm{bin}} \sum_{i=1}^{N_\mathrm{bin}} \mu_\ell (\theta_0, \phi_N^i) - \frac{1}{N_\mathrm{samp}} \sum_{j=1}^{N_\mathrm{samp}} \mu_\ell(\theta_j)
$
and variance
$
  \bar \sigma_0^2 \equiv \frac{1}{N_\mathrm{bin}^2} \sum_{i=1}^{N_\mathrm{bin}} \sigma_\ell^2 (\theta_0, \phi_N^i) + \frac{1}{N_\mathrm{samp}^2} \sum_{j=1}^{N_\mathrm{samp}} \sigma_\ell^2(\theta_j).
$ If we assume an unbiased error model, then $\bar \mu_0 = 0$.
By defining an upper limit $\sigma_{\mathrm{max},\ell}^2$ on the allowed acceptable variance in the error model for $\Delta \ell$,~\eqref{eqn:dp_sum} bounds the variance in the error in the log-posterior as $\sigma_0^2 \lesssim \sigma_{\mathrm{max},\ell}^2/N_\mathrm{bin}$, assuming $ N_\mathrm{bin} \ll N$.

\begin{figure}
\centering
\includegraphics[width=0.9\textwidth]{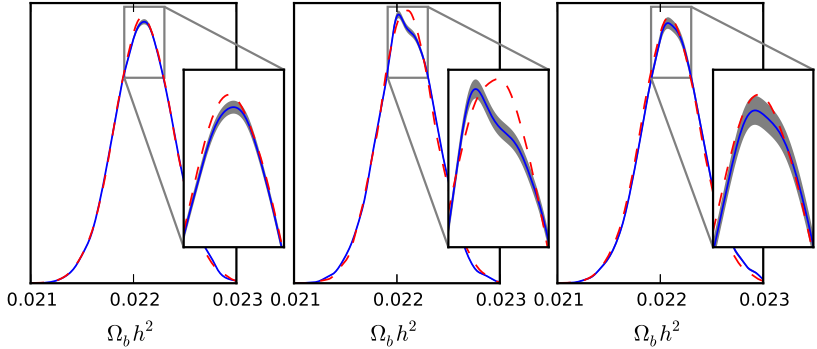}
\caption{ Posterior distribution of the baryon density $\Omega_bh^2$ with (solid blue) and without (dashed red) using emulation. The $1\,\sigma$ error of the posteriors is shown as a gray band. The left panel is for MCMC; the middle is \textsc{MultiNest}; and the right is MCMC with a larger error threshold.}
\label{fig:results}
\end{figure}

\section{Application to Cosmology}
\label{sect:results}

In~\citet{aslanyan2015learn} we demonstrated the usefulness of this approach to accelerate Monte Carlo approximation schemes in cosmology, which we review here.   We infer posterior probabilities on the six parameters of the standard $\Lambda$CDM cosmology, conditioned on cosmic microwave background (CMB) observations from the \emph{Planck} satellite.  The \emph{Planck} likelihood function~\citep{Ade:2013kta} is a combination of different components, measuring the likelihood of above-baseline fluctuations in the temperature and polarizarion of the CMB, as well as gravitational lensing effects.  To evaluate the data likelihood, one must specify a set of cosmological parameters, then forward-simulate the CMB power spectrum based on the relatively well understood physics of the plasma-dominated early universe, using packages such as CAMB~\citep{lewis2000efficient} or CLASS~\citep{blas2011cosmic}.  Each simulation takes a few seconds; and the $\Lambda$CDM posteriors are typically estimated using finely-tuned statistical samplers, such as CosmoMC~\citep{lewis2002cosmological} and MultiNest~\citep{feroz2009multinest}.

In Fig.~\ref{fig:results} we show the posterior distributions for the cosmological baryon density $\Omega_b h^2$ obtained from the non-emulated and emulated likelihoods. We have used the distance-based error model described above. We find broad agreement between the exact and approximate posteriors using two common sampling methods. We also show how the posterior error changes when increasing the allowed \emph{a priori} emulation error threshold from $-2\log \mathcal L < 0.4$ to $-2\log \mathcal L < 1.0$.  The 68\% credible regions of the marginalized posteriors for all of the $\Lambda$CDM parameters are remarkably consistent, as shown in Table~\ref{results_table}.  As compared to non-emulated methods, we found speedup factors between 67x-105x, even with  conservative tolerances on the allowed emulation error and no pre-existing training set.  Finally, using the MultiNest sampler, we report estimated marginalized likelihoods for these models as $\log Z = -4944.0 \pm 0.3$ and $\log Z_\mathrm{em} = -4944.1 \pm 0.3$,  indicating good global approximation to the non-emulated posterior.

\begin{table}
\centering
\renewcommand{\arraystretch}{1.5}
\setlength{\arraycolsep}{5pt}
\begin{eqnarray*}
\begin{array}{c|c|c|c|c|c|c}
   & \Omega_b h^2 \times 10^{2}
   & \Omega_c h^2 \times 10^{2}
   & h \times 10
   & \tau \times 10^2
   & n_s \times 10
   & \log(10^{10} A_s)
   \\
\hline
  \mathrm{Exact} & 
  2.210 \pm 0.028 &
  11.96 \pm 0.27 &
  6.80 \pm 0.12 &
  8.9 \pm 1.3 &
  9.613 \pm 0.074 &
  3.087 \pm 0.025
  \\
  \mathrm{Emul.} & 
  2.210 \pm 0.029 &
  11.96 \pm 0.27 &
  6.80 \pm 0.12 &
  8.9 \pm 1.3 &
  9.614 \pm 0.075 &
  3.088 \pm 0.025
  \\
\end{array}
\end{eqnarray*}
\caption{Parameter constraints from MCMC with and without using the emulator.}
\label{results_table}
\end{table}

\section{Conclusion}
We have demonstrated a method to accelerate posterior estimation calculations that are based on MCMC methods, either using a pre-existing training set or building one as-you-go with the sampling methods.  Our method relies on both an emulator model and a error model, which we learn separately, but the details of these models can be left up to the individual use-case.  We have demonstrated that polynomial interpolation over nearest neighbors is sufficient to replicate the results of the major $\Lambda$CDM cosmology simulations for the CMB.  Our method uses an optional, user-specified cutoff where the emulation scheme defaults to an exact evaluation of the underlying function, which is usable whenever the exact function is slow to evaluate, but not prohibitively slow.  Finally, we provide a theoretical result that shows how to calculate the error in the posterior probability for parameters of interest, conditioned on the error model; we show how to evaluate this error without any extra calls to the exact function outside of the training set.

%\subsubsection*{Acknowledgments}

\bibliography{refs}
\bibliographystyle{iclr2019_conference}

\end{document}